\documentclass[aps,prl,superscriptaddress,notitlepage,twocolumn,footinbib,floatfix]{revtex4-2}

\usepackage{xr}
\usepackage[utf8]{inputenc}
\usepackage{natbib}
\usepackage{amsmath}
\usepackage{amsfonts}
\usepackage{bm}
\usepackage{soul}
\usepackage{color} 
\usepackage[table,dvipsnames]{xcolor}
\usepackage[colorlinks,linktocpage,bookmarks=false]{hyperref}
\usepackage{booktabs}
\usepackage{ulem}

\usepackage{graphicx}
\graphicspath{{Figures/}}

\newcommand\myshade{85}
\definecolor{myrulecolor}{RGB}{150,20,0}
\colorlet{mylinkcolor}{violet}
\colorlet{mycitecolor}{YellowOrange}
\colorlet{myurlcolor}{Aquamarine}
\hypersetup{
	linkcolor  = myurlcolor!\myshade!black,
	citecolor  = mycitecolor!\myshade!black,
	urlcolor   =  myrulecolor!\myshade!black,
}

\usepackage{multirow}
\usepackage{braket}

\newcommand{\subref}[2]{\ref{#1}\hyperref[#1]{#2}}

\newcommand{\beq}{\begin{equation}}
\newcommand{\eeq}{\end{equation}}
\newcommand{\bea}{\begin{eqnarray}}
\newcommand{\eea}{\end{eqnarray}}

\newcommand{\byzo}{Ba$_3$Yb$_2$Zn$_5$O$_{11}$}
\newcommand{\blzo}{Ba$_3$Lu$_2$Zn$_5$O$_{11}$}
\newcommand{\musr}{$\mu^+$SR}

\newcommand{\meV}{\ {\rm meV}}

\renewcommand\[{\begin{equation}}
\renewcommand\]{\end{equation}}
\renewcommand{\emph}[1]{\textit{#1}}

\begin{document}

\title{Beyond Single Tetrahedron Physics of Breathing Pyrochlore Compound \byzo{}}

\author{Rabindranath Bag}
\thanks{equal contribution}
\affiliation{Department of Physics, Duke University, Durham, NC 27708, USA}
\author{Sachith E. Dissanayake}
\thanks{equal contribution}
\affiliation{Department of Physics, Duke University, Durham, NC 27708, USA}
\author{Han Yan}
\affiliation{Rice Academy of Fellows, Rice University, Houston, TX 77005, USA}
\author{Zhenzhong Shi}
\affiliation{Department of Physics, Duke University, Durham, NC 27708, USA}
\author{David Graf}
\affiliation{National High Magnetic Field Laboratory and Department of Physics, Florida State University, Tallahassee, Florida 32310, USA.}
\author{Eun Sang Choi}
\affiliation{National High Magnetic Field Laboratory and Department of Physics, Florida State University, Tallahassee, Florida 32310, USA.}
\author{Casey Marjerrison}
\affiliation{Department of Physics, Duke University, Durham, NC 27708, USA}
\author{Franz Lang}
\affiliation{Clarendon Laboratory \& Physics Department, University of Oxford, Parks Road, Oxford OX1 3PU, United Kingdom}
\author{Tom Lancaster}
\affiliation{Department of Physics, Centre for Materials Physics, Durham University, Durham DH1 3LE, United Kingdom}
\author{Yiming Qiu}
\affiliation{NIST Center for Neutron Research, National Institute of Standards and Technology, Gaithersburg, Maryland 20899, USA}
\author{Wangchun Chen}
\affiliation{NIST Center for Neutron Research, National Institute of Standards and Technology, Gaithersburg, Maryland 20899, USA}
\author{Stephen J. Blundell}
\affiliation{Clarendon Laboratory \& Physics Department, University of Oxford, Parks Road, Oxford OX1 3PU, United Kingdom}
\author{Andriy H. Nevidomskyy}
\affiliation{Department of Physics and Astronomy, Rice University, Houston, TX 77005, USA}
\author{Sara Haravifard}
\email{sara.haravifard@duke.edu}
\affiliation{Department of Physics, Duke University, Durham, NC 27708, USA}
\affiliation{Department of Materials Sciences and Mechanical Engineering, Duke University, Durham, NC 27708, USA}

\date{\today}

\begin{abstract}
    Recently a new class of quantum magnets, the so-called breathing pyrochlore spin systems, have attracted much attention due to their potential to host exotic emergent phenomena. Here, we present magnetometry, heat capacity, thermal conductivity, Muon-spin relaxation, and polarized inelastic neutron scattering measurements performed on high-quality single-crystal samples of breathing pyrochlore compound \byzo{}. We interpret these results using a simplified toy model and provide a new insight into the low-energy physics of this system beyond the single-tetrahedron physics proposed previously.
\end{abstract}

\maketitle
 \vspace{-5mm}

\begin{figure}[t!]
	\centering 
	\includegraphics[width=\columnwidth]{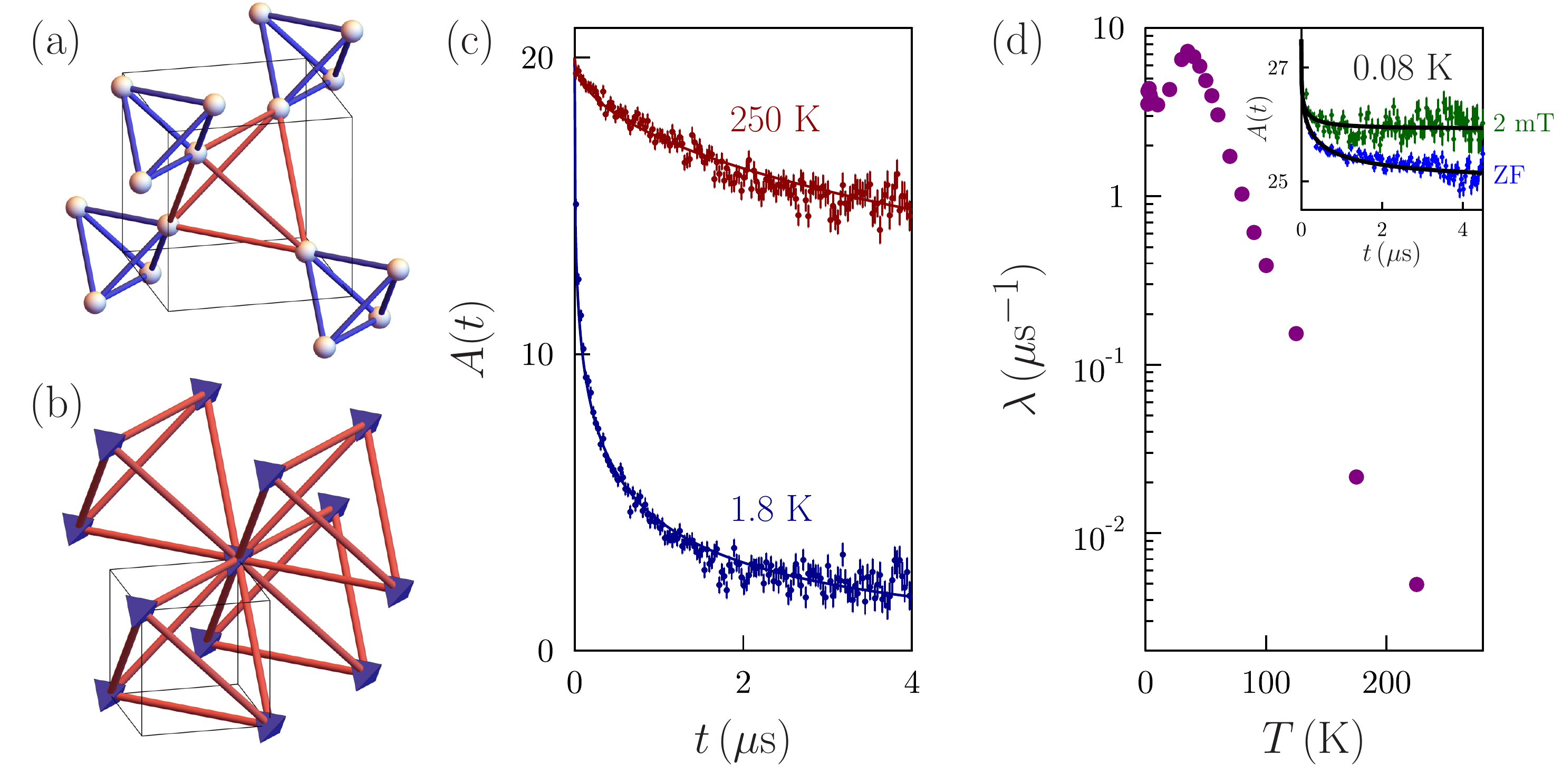}
	\caption{
	(a) Structure of the breathing pyrochlore lattice.
	Considering each $A$-tetrahedron (blue) as a single site and retaining the bonds on the $B$-tetrahedron (red), we obtain a  face-centered cubic (FCC) lattice shown in (b). The cube in solid black line marks the repeating unit in the left panel in the FCC lattice. (c) Zero-field muon asymmetry for two temperatures. Solid lines represent fits to  $A(t)=A(0)e^{-(\lambda t)^\beta}$ with $A(0)$ fixed across all temperatures. (d) The temperature dependence of the fitted relaxation rate $\lambda$.  The inset shows ultra-low temperature data, demonstrating the absence of long range order at 0.08~K and the effect of applying a small 2~mT longitudinal field.
	    	\label{fig:Pyro_FCC_musr}
	} 

\end{figure}

Frustrated quantum magnets provide a fruitful arena to search for novel quantum phenomena \cite{lacroix2011introduction,Balents2010}. Pyrochlore lattice magnets, in which magnetic ions form corner-sharing regular tetrahedra, are one of the most studied frustrated systems in three-dimension \cite{SUBRAMANIAN1983,GardnerRevModPhys,RauGingrasAnnuRevCMP,HallasXYAnnuRevCMP}. In the pyrochlore system the conventional magnetic ordering is suppressed by the geometrically frustrated lattice, consequently resulting in emergence of exotic phases \cite{GaoNatPhys2019,Savary_Er2Ti2O7,KimuraQuantum2013,OitmaaPRB2013,BramwellPRL2001,Ramirez1999,MirebeauPRL2004,ChangNaturecomm2012,ZhitomirskyPRL2012,RauPRB2016,RossPRX2011,HayrePRB2013,ThompsonPRL2011}. Recently a new class of systems, the so-called breathing pyrochlore magnets, have attracted much attention due to their potential to host exotic phenomena and topological phases \cite{SavaryPRB2016,li2016weyl,aoyama2021effects,aoyama2021hedgehog}. In breathing pyrochlore compounds the lattice inversion symmetry at each site is broken due to the different sizes of up-pointing and down-pointing tetrahedra, thus resulting in large Dzyaloshinskii-Moriya (DM) interactions on the two tetrahedra \cite{talanov2020formation} (see Fig.~\subref{fig:Pyro_FCC_musr}{(a,b)} for the structure of breathing pyrochlores).
On the theory front, recent  works have shown that breathing pyrochlore spin systems can host novel physics including classical rank-2 $U(1)$ spin-liquid states \cite{Yan2020PhysRevLett}, quantum fractons \cite{HanPRB2022}, competing quantum spin liquids \cite {chern2022competing}, and hedgehog lattices of magnetic monopoles and antimonopoles \cite{aoyama2021hedgehog}. Thus, it is of great interest to synthesize and understand breathing pyrochlore materials. The majority of the work performed on the breathing pyrochlore-based compounds have focused on Cr-based spinels with $S=3/2$ \cite {GenPRB2020,LeePRB2016,OkamotoPRL2013,OkamotoJPSJ2015,PokharelPRB2018,GhoshNPJQM2019, gao2021hierarchical,gui2021ferromagnetic,sharma2022synthesis}, while the studies performed on quantum systems with $S=1/2$ remain limited to \byzo{} in powder form \cite{KimuraPRB2014,RauPRL2016,HakuPRB2016,HakuJPSJ2016,haku2017neutron,RauJPCM_2018}. Our recent work reported the first comprehensive neutron scattering studies performed on single-crystal sample of \byzo{} \cite{Dissanayake2021}. 

\vspace{-1mm}
We successfully grew single crystal samples of breathing pyrochlore \byzo{} using the modified optical floating zone technique \cite{Dissanayake2021}. Inelastic neutron scattering studies using our single crystal sample revealed that the single-tetrahedron model with isolated tetrahedra can explain the high-temperature and high-energy regime of the collected data~\cite{Dissanayake2021}. However, the diffuse neutron scattering performed at low-temperature and low-energy reveals features which cannot be understood with this model \cite{Dissanayake2021}. Pair distribution function (PDF) analyses performed on high quality powder neutron diffraction data provided evidence for the absence of chemical disorder within experimental resolution. Single crystal X-ray diffraction studies also found no evidence of site disorder~\cite{Dissanayake2021}. Diffuse neutron scattering on single crystal samples~\cite{Dissanayake2021}, and the low temperature heat capacity data collected on powder samples \cite{HakuPRB2016,RauJPCM_2018} suggest that physics beyond the single tetrahedron, with small but finite inter-tetrahedron interactions, is essential to capture the behavior of this quantum breathing pyrochlore system. This calls for additional experimental probes at low temperature with higher resolution, so that we can study the subtle changes in magnetic properties of \byzo{} with higher accuracy.

\vspace{-1mm}
In this letter, we report low-temperature heat capacity measurements in applied field, ultra-sensitive magnetic susceptibility, thermal conductivity, muon spin relaxation (\musr), and polarized inelastic neutron scattering measurements of the ytterbium based breathing pyrochlore compound \byzo{} in single-crystalline form, to investigate the intrinsic low temperature magnetic properties and provide a first look into the physics governing the low-energy regime of this system.  We propose a simplified model that captures the field dependence of the heat capacity for lower field region well and provides a scenario beyond the previously reported single-tetrahedron physics, with finite inter-tetrahedron coupling necessary to interpret the experimental results.

In order to search for any trace of magnetic order, we performed \musr{} measurements using a large powder sample of \byzo{} on the GPS spectrometer at the Swiss Muon Source at PSI, and also using co-aligned single crystal samples of \byzo{} mounted in a dilution refrigerator at the MuSR spectrometer of the ISIS Muon Source. In order to understand the origin of the contributions to the \musr{} signal, we carried out density functional theory (DFT) calculations to locate the most probable muon stopping sites, and assess the degree of perturbation the muon-probe causes in the material \cite{moller2013playing}, which we find to be small in this system (for details see \cite{supplementary}). The measured muon asymmetry $A(t)$ in zero field (ZF) exhibits a non-oscillating but relaxing time dependence [see Fig.~\subref{fig:Pyro_FCC_musr}{(c)}], the relaxation rate of which decreases on warming above around 50~K.  This relaxation can be fitted to a stretched exponential form $A(t)=A(0)e^{-(\lambda t)^\beta}$ where the exponent $\beta\approx 0.33$, indicating that the spin fluctuations that give rise to the relaxation have a range of timescales (an exponent of $\beta=1$ would indicate a single fluctuation time, and exponents less than one are often found in systems with complex spin dynamics, see e.g.~\cite{Keren2001}).  The temperature dependence of $\lambda$ is plotted in Fig.~\subref{fig:Pyro_FCC_musr}{(d)}, illustrating the decrease in $\lambda$ with increasing temperature at high temperatures, corresponding to the thermally-induced increase in spin-fluctuation rate ($\lambda \propto \nu^{-1}$ in the fast-fluctuation limit, where $\nu$ is a characteristic fluctuation rate \cite{Blundell2022}). There is a maximum in $\lambda$ around 35~K, the low-temperature approach to which we attribute to the switching on of excitations associated with the intra-tetrahedral interactions (note that the energy gap of 0.38~meV \cite{HakuJPSJ2016} to the next crystal field level is larger than any thermal energy in this experiment).  Our experiments at milliKelvin temperatures were hampered by the very small size of the single crystals (leading to a small relaxation amplitude) but nevertheless demonstrate that the relaxing signal persists down to 0.08~K [see inset to Fig.~\subref{fig:Pyro_FCC_musr}{(d)}], with no appearance of an oscillatory signal, ruling out the development of long range magnetic order. It is noteworthy to add that the ultra-low temperature persistent slow dynamics are consistent with what is seen in many other frustrated systems, and in this case one can argue that they might be connected with the very weak inter-tetrahedra interactions. We will explore such possibility in more details later as we discuss the ultra-low temperature thermodynamics results. 

\begin{figure*}[ht]
	\centering 
	\includegraphics[width=1\textwidth]{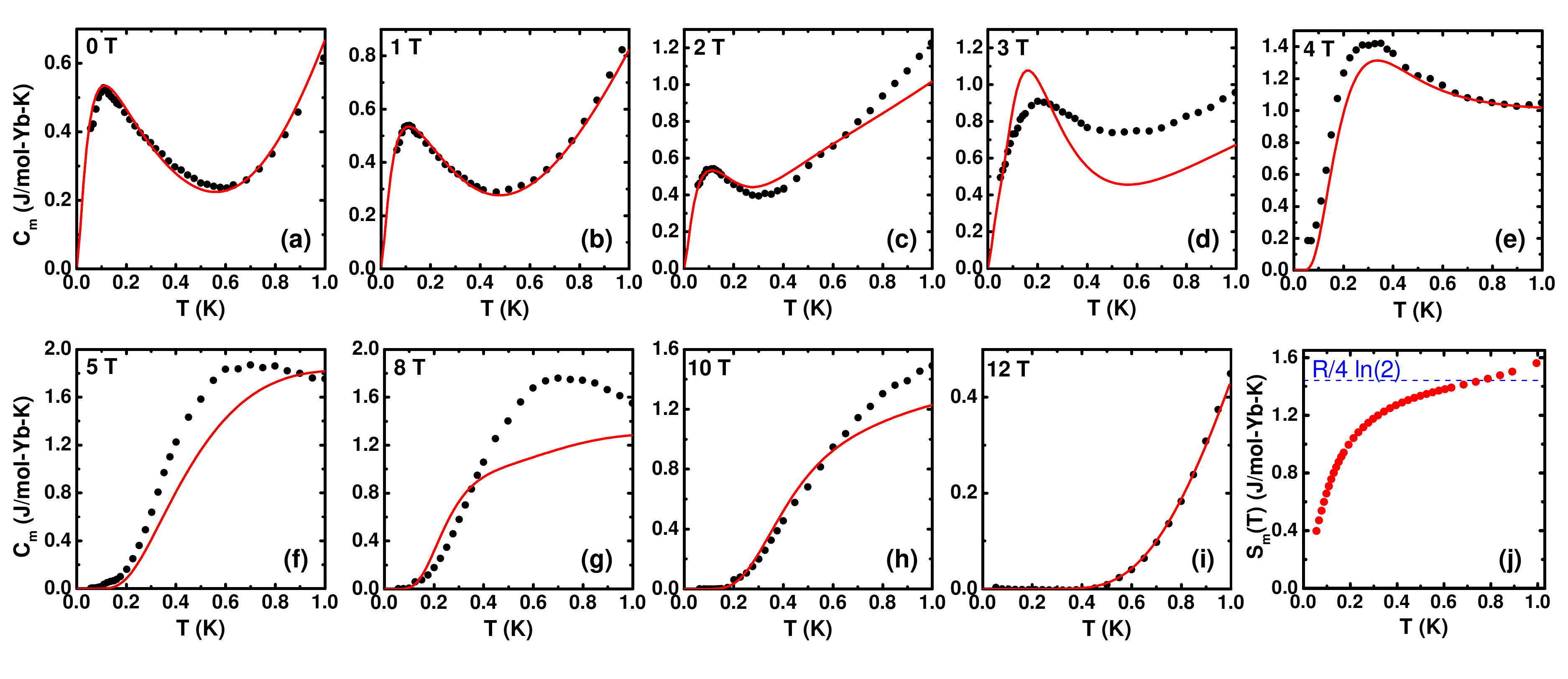}
  \vspace{-3mm}
	\caption{
	(a)-(i) Low temperature magnetic specific heat (C$_m$) of \byzo{} single crystal samples for different fields from 0 T to 12 T. The phonon contribution was subtracted using the  iso-structural, non-magnetic compound \blzo{}. Red lines are fits based on the simplified model outlined in the text \cite{supplementary}.
	(j) Magnetic entropy of \byzo{} at zero magnetic field. 
	} 
	\label{fig:HC}
\end{figure*}

We show in Fig.~\subref{fig:HC}{(a-i)} the magnetic heat capacity for a \byzo{} single crystal sample from 54 mK to 1 K under different applied magnetic fields, with the phonon contribution subtracted using the results of measurements made on iso-structural, non-magnetic \blzo.The magnetic entropy at zero field is shown in Fig.~\subref{fig:HC}{(j)}. The heat capacity data were collected on two different \byzo{} single-crystal samples (grown using different techniques) and are compared with the reported powder \byzo{} sample \cite{supplementary}.  Previous reports discussed the possibility of having defects, such as structural disorder, as an underlying cause for the peak observed at low temperatures, whereas our heat capacity data collected on multiple single crystal samples excludes the existence of measurable defect effects such as structural disorder. To further elaborate on this, we show in Fig.~\ref{fig:HC_supp} of the Supplementary Materials \cite{supplementary} the results obtained for two single-crystal samples grown with different techniques (sample 1 and sample 2). The peak position at 110 mK remains the same for both single-crystal samples and agrees with the reported powder study by Haku \textit{et al.} \cite{HakuPRB2016}. 
This is while the fit to the data is significantly improved using the model we employed to analyze the data. We explain the details of this model in the following. Additionally, here we show the field-dependence of the low-temperature feature which agrees reasonably with our proposed model, in particular for the low and high field region.

As shown in Fig.~\subref{fig:HC}{(j)}, there is $\frac{1}{4}R\ln2$ entropy release per Yb ion, corresponding to an effective pseudo-spin-1/2 degree of freedom on each tetrahedron. As discussed by Rau \textit{et al.} \cite{RauJPCM_2018} the experimental specific heat results collected at $T<0.4$ K disagree with the single tetrahedron model, leading us, to propose that this release of entropy is related to the inter-tetrahedron interactions. This is because in the single tetrahedron theory, the two lowest states are robustly degenerate, and the third state lies much higher in energy (at $ \sim 0.5~\text{meV}$). Although for a finite external magnetic field, the degeneracy of the two lowest states is expected to be lifted, the energy splitting is much smaller than  $\sim0.01~\text{meV}$, which cannot explain the  broad peak in heat capacity measurement in Figs.~\subref{fig:HC}{(a,b)}.

The above observations suggest that the low-energy properties of \byzo{} cannot be explained by the single tetrahedron theory, even if tuning the exchange parameters is allowed. Instead the specific heat data can be  understood quantitatively by introducing inter-tetrahedron interactions. To this end, we have constructed an effective low-energy model, regarding the two nearly degenerate lowest energy states (out of 16) on the $A$ tetrahedra as a pseudo-spin $\frac{1}{2}$. This is justified by the fact that the other states lie at much higher energies ($E_3>0.3$~meV)~\cite{RauJPCM_2018} relative to the range $T<1$~K in our specific heat data.

From exact diagonalization on a single tetrahedron, we can determine the wave-function of the two lowest states exactly, which form the two-dimensional ($E$) irreducible representation of the $T_d$ point group. In the limit of vanishing intra-tetrahedron DM interaction, these states span the two-dimensional Hilbert space of two-dimer coverings of the four sites. Note that there are three such possible dimer coverings classically, but one of them is linearly dependent of the other two. The small DM interaction in $A$-tetrahedra tunes the wave-function away from the perfect dimer-covering states \cite{RauPRL2016,Curnoe2007PhysRevB},
but does not lift their degeneracy, justifying our treating them as pseudo-spin $\frac{1}{2}$ degrees of freedom. We then consider the interaction between these pseudo-spins via the weak bonds on the $B$-tetrahedra [shown in red in Fig.~\subref{fig:Pyro_FCC_musr}{(a,b)}]. Shrinking every $A$-tetrahedron to a point connected by these weak bonds, the effective low-energy model becomes an face-centered cubic (FCC) lattice of pseudo-spins with nearest neighbor interactions [c.f.~Fig.~\subref{fig:Pyro_FCC_musr}{(b)}]. The effective interactions are  the original interactions between the physical spins projected onto the pseudo-spin Hilbert space. 

\begin{figure*}[tb!]
	\centering
	\includegraphics[width=1\textwidth]{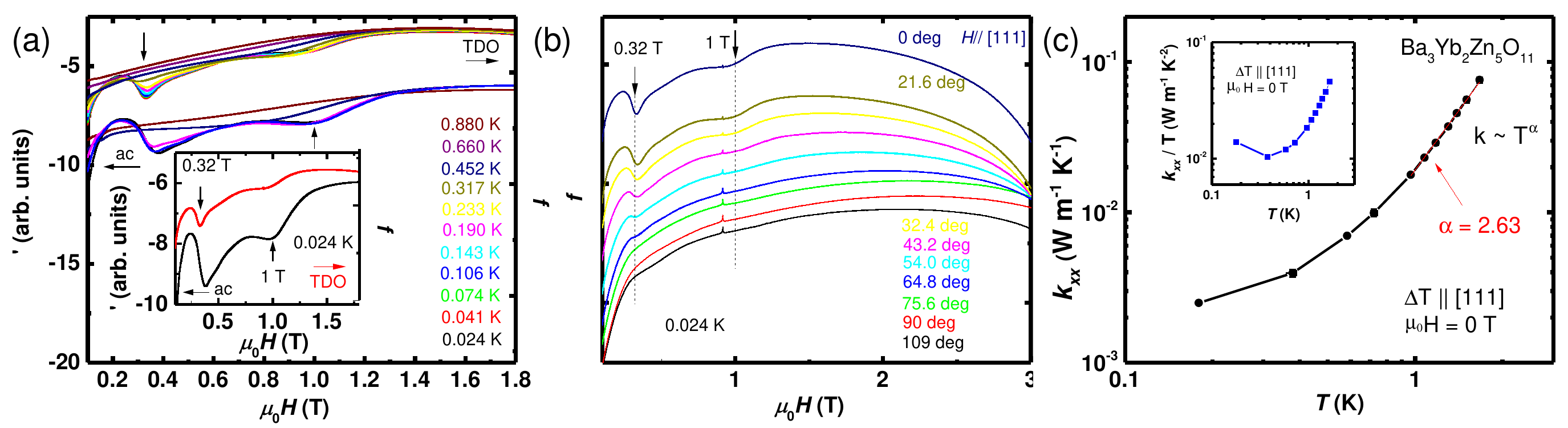}
	\caption{Low-$H$ anomalies beyond the single-tetrahedra model: (a) magnetic ac-susceptibility $\chi'$ (left axis) and TDO frequency ($f$) (right axis) as a function of external magnetic field ($H\parallel [111]$) at different temperatures. Two anomalies are marked by arrows. $\chi'$ is measured with an oscillating frequency of 1616 Hz. Measurements ($\chi'$) with different frequency (87.1 Hz) reproduce the two anomalies at the same fields. Inset shows zoomed-in plot of $\chi'$ (left axis) and TDO $f$ (right axis) at $T =$ 0.024 K. (b) TDO frequency ($f$) data is shown for different rotation angles between the field and the crystal $[111]$ axis at 0.024 K. Traces are shifted vertically for clarity. (c) Thermal conductivity data ($\kappa_{xx}$ vs $T$) for \byzo{} at $\mu_0H$ = 0 T; $\delta T$ $||$ $[111]$. Solid red lines are power-law fit to $\kappa_{xx}$ data. A clear saturation of $\kappa_{xx}$ at low $T$ is seen for \byzo. Inset shows $\kappa_{xx}$/T vs $T$ plot. 
		\label{fig:TDO_AC_LowField}
	}
\end{figure*}

We consider the simplest model of the effective  interactions between neighboring $A$-tetrahedra pseudo-spins:
\[
H=J_{xy}(s^x_is^x_j +s^y_is^y_j) + J_zs^z_is^z_j 
\label{eq:ham}
\]
We find that choosing ferromagnetic $J_z =-0.005$~meV and antiferromagnetic $J_{xy}= 0.0125~\text{meV}$ in this XXZ model 
can reproduce the zero-field specific heat very well [Fig~\subref{fig:HC}{(a)}] (for details, see Supplementary Material \cite{supplementary}). The origin of the $110$~mK peak in the specific heat is due to the ferromagnetic ordering of the pseudo-spins, which spontaneously lifts the two-fold degeneracy of the single-tetrahedron model. Linear spin-wave theory then produces pseudo-magnons of the bandwidth $\sim J_{xy}$ which propagate on the FCC lattice, as depicted in Fig.~\ref{fig:FCC_SWDOS}. It is important to note that pseudo-magnons are not conventional spin waves, but rather collective excitations of the dimer-covering states spanned by pseudo-spin degrees of freedom on $A$-tetrahedra, and hence may be challenging to detect by inelastic neutron scattering.

The effective low-energy model in Eq.~\eqref{eq:ham} is expected to work in a moderately large applied magnetic field, provided its strength  does not exceed the energy gap ($E_3\sim 0.38$~\meV) to the first excited state beyond the $E$-doublet  in the single-tetrahedron model~\cite{RauJPCM_2018} . The magnetic field splits pseudo-spin degrees of freedom in Eq.~\eqref{eq:ham} at an energy scale much smaller than the parameters in the effective model (Fig.~S3 in  \cite{supplementary}).
The main effect of the field, from the exact diagonalization of a single tetrahedron, is to shift the higher-energy states downwards, which we treat as flat bands. This approximation breaks down at a critical value of the field $B_c\sim4$~T when the lowest excited state $E_3$ crosses the ground state doublet, resulting in a phase transition. Our model \eqref{eq:ham} does not apply in this regime or higher fields. In in the limit of high fields $B>10$~T,  we are able to obtain a good match with the  experimental specific heat  [see Fig.~\subref{fig:HC}{(i)}] by using a single tetrahedron theory, which predicts a unique non-degenerate ground state separated by a large gap from the higher-lying states. For intermediate field strengths, one cannot ignore the effect of the excited states, which result in the ground state level crossing as already noted. The minimal model then becomes rather complicated, with a vast range of unknown parameters, whose determination lies beyond the scope of the present work.

Further insight into the effect of weak magnetic fields can be gleaned from the magnetic ac-susceptibility, which we measured on two separate \byzo{} crystals. The results are shown in Fig.~\subref{fig:TDO_AC_LowField}{(a)}, which show two anomalies at  $\mu_0 H_{c1}=0.32$~T and $\mu_0H_{c2}=1.0$~T upon cooling at low temperatures $T \simeq0.3$~K. The two anomalies are also seen in the tunnel diode oscillator (TDO) measurements on the same crystals and do not shift appreciably with field when the crystal is rotated away from $[111]$ orientation [see Fig.~\subref{fig:TDO_AC_LowField}{(b)}].
These anomalies are independent of two different oscillation frequencies (1616 Hz and 87.1 Hz) of the ac-susceptibility measurement, suggesting the signal is unrelated to spin freezing and consistent with the \musr{} study showing absence thereof. The most likely explanation for the anomalies is the level crossing at the corresponding fields $\mu_0 H_{c1}$  and $\mu_0 H_{c2}$. Given the high frustration of the FCC lattice, it is possible for the system to go through several different phases. The exact phases and phase transitions cannot be determined by current experimental data, and await future effort. Importantly, the explanation must involve inter-tetrahedron couplings, because the single tetrahedron model would predict a nonmagnetic $S_{\mathrm{eff}}$ = 0 ground state at low temperatures ($\leq 0.5$~K) and thus a featureless magnetic susceptibility, contrary to what we see in our measurements.

In order to further understand the nature of the low-lying states observed in the heat capacity and magnetometry data, low temperature thermal conductivity measurements were carried out on single crystal sample of \byzo{} \cite{supplementary}. A power law fit ($\kappa_{xx} \sim T^\alpha$) is performed on the collected data at the high-$T$ region and the value of the exponent ($\alpha$) is found to be 2.63. For a nonmagnetic insulator, $\kappa_{xx}$ at very low temperature is only due to the contribution from phonons \cite{XuPRL2016,SutherlandPRB2003}. However, the exponent value obtained for \byzo{} reveals additional contributions coming from various quasi-particles such as phonons, spinons and magnons, as well as different scattering channels for the heat current~\cite{RaoNaturecomm2021,SutherlandPRB2003,LiPRB2008,NiNatureComm2021}. Interestingly, at low $T$, a clear saturation of $\kappa_{xx}$ in \byzo{} is seen. Such saturation is not expected in conventional magnets where both phonons and magnons freeze out, but suggest itinerant fermionic (spinon) excitations expected in gapless spin liquids candidates \cite{XuPRL2016,YuPRL2018,HopePRX2019,NiNatureComm2021}. Further studies are needed to probe the nature of these low-$T$ magnetic excitations.

\begin{figure}[t!]
	\centering 
	\includegraphics[width=\columnwidth]{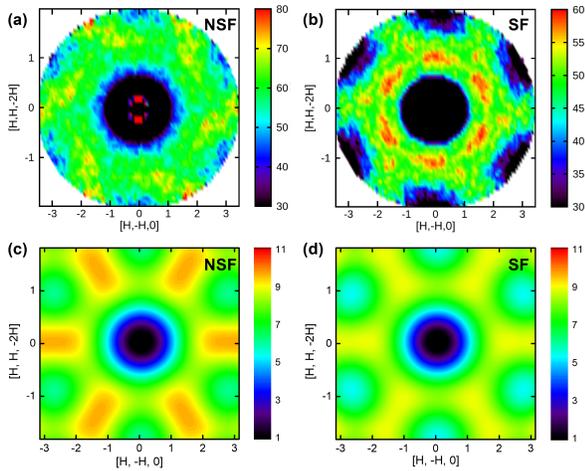}
	\caption{ (a) and (b) show the polarized neutron scattering data measured at $T=0.1$ K and $E=0.5$ meV, after subtracting data at $T=50$ K and $E=0.5$ meV as the background. (c) and (d) present the calculations of NSF and SF scattering using single tetrahedron theory outlined in the text. 
	} 
	\label{fig:BYZO_Polarized_Figure}
\end{figure}
To provide a better understanding for the underlying physics governing the physics of \byzo{} at ultra-low temperatures, we have complemented these efforts by performing polarized inelastic neutron scattering measurements on single crystal sample (for details see \cite{supplementary}). The results of this experiment is shown in Figs.~\subref{fig:BYZO_Polarized_Figure}{(a,b)}, in which $T=0.1$ K and $E=0.5$ meV. Non-spin-flip (NSF) and spin-flip (SF) scattering plots represent the spin dynamics along $[111]$ cubic direction and the $[{h+k},-{h}+k,-2k]$  plane, respectively. We show in Figs.~\subref{fig:BYZO_Polarized_Figure}{(c,d)} the corresponding calculations for NSF and SF scattering using single tetrahedron theory. The experimental data and the single tetrahedron calculations for the NSF case are in qualitative agreement, whereas the comparison of the SF results shows a distinct difference - in particular at low-Q range. Considering that the energy resolution for our experiment at elastic channel was about 0.3 meV, one can argue that the experimental data captures excitations as low as 0.1 meV. Thus, the observed difference between the single tetrahedron theory calculations and the SF experimental result could possibly be due to the weak inter-tetrahedron interactions. This argument would be aligned with what we discussed earlier to explain the thermodynamics results at ultra-low temperatures. Further experiments with higher energy resolution and corresponding calculations are needed to confirm this theory.

In conclusion, we report the evidence of \byzo{} being the first interacting quantum breathing pyrochlore spin-1/2 system, contrary to previous theoretical treatments assuming uncorrelated tetrahedra. The ultra-low temperature \musr{} results demonstrate presistent spin dynamics, and the thermal conductivity data, collected at the same temperature range, suggest itinerant fermionic excitations. Furthermore, the polarized inelastic neutron scattering results collected at low-energy range appears to divert from the previously proposed single tetrahedron model. Clearly, follow up experiments are needed to confirm the nature of the spin interactions in \byzo{} at low-temperature/low-energy regime, however, based on our current experimental results we suggest that a simple effective XXZ  model, formulated in terms of the lowest-energy doublets per tetrahedron, can account for the ultra-low temperature specific heat measured at low external field $\mu_0H\lesssim 1~\text{T}$. Additionally, in this work we report that the ac-susceptibility and TDO measurements show two anomalies at $\mu_0H=0.32, 1.0~\text{T}$, suggesting that the system  goes through two transitions yet to be understood. Our findings open the gates toward a landscape of breathing pyrochlore materials and non-trivial exotic phases of matter, including fracton physics, that can be realized within \cite{Yan2020PhysRevLett,HanPRB2022,chern2022competing}.

\begin{acknowledgments}
We are thankful to William Steinhardt, Jeffrey Rau, and Michel Gingras for fruitful discussions. Research performed at Duke University is supported by the National Science Foundation grant no. DMR-1828348. A portion of this work was performed at the National High Magnetic Field Laboratory, which is supported by the National Science Foundation Cooperative Agreement No. DMR-1157490 and DMR-1644779, the State of Florida and the U.S. Department of Energy. We acknowledge the support of EPSRC (UK) and the facilities of the Hamilton
HPC Service of Durham University. Data from the UK effort will be made available via XXX. For the purpose of open access, the authors have applied a Creative Commons Attribution (CC BY) license to any Author Accepted Manuscript version arising. Muon measurements were made at the Swiss Muon Source and the STFC-ISIS Facility and we are grateful for the provision of beamtime. Access to MACS polarized neutron measurements was provided by the Center for High Resolution Neutron Scattering, a partnership between the National Institute of Standards and Technology and the National Science Foundation under Agreement No. DMR-1508249. A.H.N. and H.Y. were  supported by the National Science Foundation grant no. DMR-1917511. S.H. acknowledges support provided by funding from William M. Fairbank Chair in Physics at Duke University, and from the Powe Junior Faculty Enhancement Award.

\end{acknowledgments}

\bibliography{main}

\end{document}


\title{Supplementary Material for ``Beyond Single Tetrahedron Physics of Breathing Pyrochlore Compound \byzo{}''}

\author{Rabindranath Bag}
\thanks{equal contribution}
\affiliation{Department of Physics, Duke University, Durham, NC 27708, USA}
\author{Sachith E. Dissanayake}
\thanks{equal contribution}
\affiliation{Department of Physics, Duke University, Durham, NC 27708, USA}
\author{Han Yan}
\affiliation{Rice Academy of Fellows, Rice University, Houston, TX 77005, USA}
\author{Zhenzhong Shi}
\affiliation{Department of Physics, Duke University, Durham, NC 27708, USA}
\author{David Graf}
\affiliation{National High Magnetic Field Laboratory and Department of Physics, Florida State University, Tallahassee, Florida 32310, USA.}
\author{Eun Sang Choi}
\affiliation{National High Magnetic Field Laboratory and Department of Physics, Florida State University, Tallahassee, Florida 32310, USA.}
\author{Casey Marjerrison}
\affiliation{Department of Physics, Duke University, Durham, NC 27708, USA}
\author{Franz Lang}
\affiliation{Clarendon Laboratory \& Physics Department, University of Oxford, Parks Road, Oxford OX1 3PU, United Kingdom}
\author{Tom Lancaster}
\affiliation{Department of Physics, Centre for Materials Physics, Durham University, Durham DH1 3LE, United Kingdom}
\author{Yiming Qiu}
\affiliation{NIST Center for Neutron Research, National Institute of Standards and Technology, Gaithersburg, Maryland 20899, USA}
\author{Wangchun Chen}
\affiliation{NIST Center for Neutron Research, National Institute of Standards and Technology, Gaithersburg, Maryland 20899, USA}
\author{Stephen J. Blundell}
\affiliation{Clarendon Laboratory \& Physics Department, University of Oxford, Parks Road, Oxford OX1 3PU, United Kingdom}
\author{Andriy H. Nevidomskyy}
\affiliation{Department of Physics and Astronomy, Rice University, Houston, TX 77005, USA}
\author{Sara Haravifard}
\email{sara.haravifard@duke.edu}
\affiliation{Department of Physics, Duke University, Durham, NC 27708, USA}
\affiliation{Department of Materials Sciences and Mechanical Engineering, Duke University, Durham, NC 27708, USA}

\date{today}
\maketitle

\section{Experimental Methods}

\label{sec:Methods}

\subsection{Sample Synthesis and Single Crystal Growth}
Polycrystalline samples of \byzo{} were synthesized in a standard solid state reaction route. Starting precursors of Yb$_2$O$_3$, ZnO, and BaCO$_3$ were used in a stoichiometric ratio of 1:5:3 and were finely mixed using a mortar and a pestle. The mixture was finally sintered at 1140° C inside a box furnace for 48 hours with several intermediate grindings. The phase purity is confirmed using powder X-ray diffraction (PXRD). The powder was compressed into cylindrical feed and seed rods using a hydrostatic press of 700 bar pressure. To increase the density, the feed and seed rods were sintered at 1150° C inside a vertical Bridgman tube furnace. The centimeter sized large and high quality \byzo{} single crystals were grown using a four mirror optical floating zone furnace with xenon lamps. The \byzo{} crystals were grown with slightly different growth conditions: crystal sample 1 and crystal sample 2 were grown in presence of O$_2$ pressure of 0.6 Mpa and 0.9 MPa, with a growth speed of 10 mm/h and 6 mm/h, respectively. To get a homogeneous mixture at the liquid zone, feed and seed were rotated in opposite direction with a rotation speed of 10-15 rpm during the growth process. Low temperature heat capacity and thermal conductivity measurements were carried out on two \byzo{} crystals (crystal sample 1 and crystal sample 2) grown under different growth condition as explained. The data were found to be reproducible using two crystal samples 1 and 2. 

\subsection{Muon Spin Relaxation}
\label{sec:muSr}
Muon Spin Relaxation (\musr{}) data were collected on (i) a powder sample using the GPS spectrometer at the Swiss Muon Source, located at the Paul Scherrer Institute in Switzerland and (ii) four coaligned \byzo{} single crystal samples using the MuSR spectrometer at the ISIS Muon source, Rutherford Appleton Laboratory, UK. The sample at PSI was measured in a $^4$He flow cryostat in zero-field (ZF).  The experiment at ISIS was performed using  a dilution refrigerator, with data collected both in zero-field (ZF) and longitudinal field (LF). The mass of the sample in the ISIS experiment was small and so about 90\% of the signal was background; in contrast, the PSI experiment had almost complete stopping of the muons in the sample, but only reached down to 1.8~K.  However, the ISIS data were able to show that there is very little temperature dependence between 2~K and the base temperature of the dilution refrigerator (0.08~K).
The muon site calculations were carried out using the MuFinder software \cite{huddart2022mufinder} and the plane-wave-based code Castep \cite{clark2005first} using the local density approximation. Muons, modelled by an ultrasoft hydrogen pseudopotential, were initialized in range of low-symmetry positions and the structure was allowed to relax (keeping the unit cell fixed) until the change in energy per ion was less than $1 \times10^{-5}$~eV. We used a cutoff energy of 544~eV and a $1 \times 1 \times 1$ Monkhurst-Pack grid \cite{monkhorst1976special} for k-point sampling.
The lowest-energy muon site is found to be $\approx 1$~\AA\ from an oxygen atom, outside a small Yb$^{3+}$ tetrahedron. It causes a modest distortion, with the nearest- neighbour oxygen atoms shifted towards the muon site by $\approx 0.18$~\AA, with the next-nearest neighbour oxygen shifted by $\approx 0.15$~\AA\ towards the muon. The nearest Yb$^{3+}$ is pushed away from the muon by $\approx 0.15$~\AA, such that the smallest $\mu^+$--Yb distance is 2.4~\AA. Other muon sites were found at slightly higher energy ($\approx$0.37~eV above the lowest-energy sites) that might also be realized. These again involve the muons localizing near an oxygen, but this time well separated from the Yb$^{3+}$ ions (with the distance to the nearest Yb ion being $\approx 4$~\AA), such that the next-nearest neighbor is a Zn (separated by 2.1~\AA) pushed around 0.6~\AA\ away from the muon. Owing to the distances involved, these more distorting sites are expected to be less well-coupled to the local magnetism and so, even if they are realized, they will have less influence on the muon spectra.  More importantly, Yb$^{3+}$ is a Kramers ion and so any muon-induced distortion is not going to result in the splitting of a crystal field ground-state doublet (in contrast to certain Pr-containing compounds for example \cite{Foronda2015}), and so its effect on the magnetic properties of the system is likely to be minimal, particularly as the gap between the crystal field ground state and any excited state is very large \cite{HakuJPSJ2016}.

\begin{figure}[ht]
	\centering 
	\includegraphics[width=1\columnwidth]{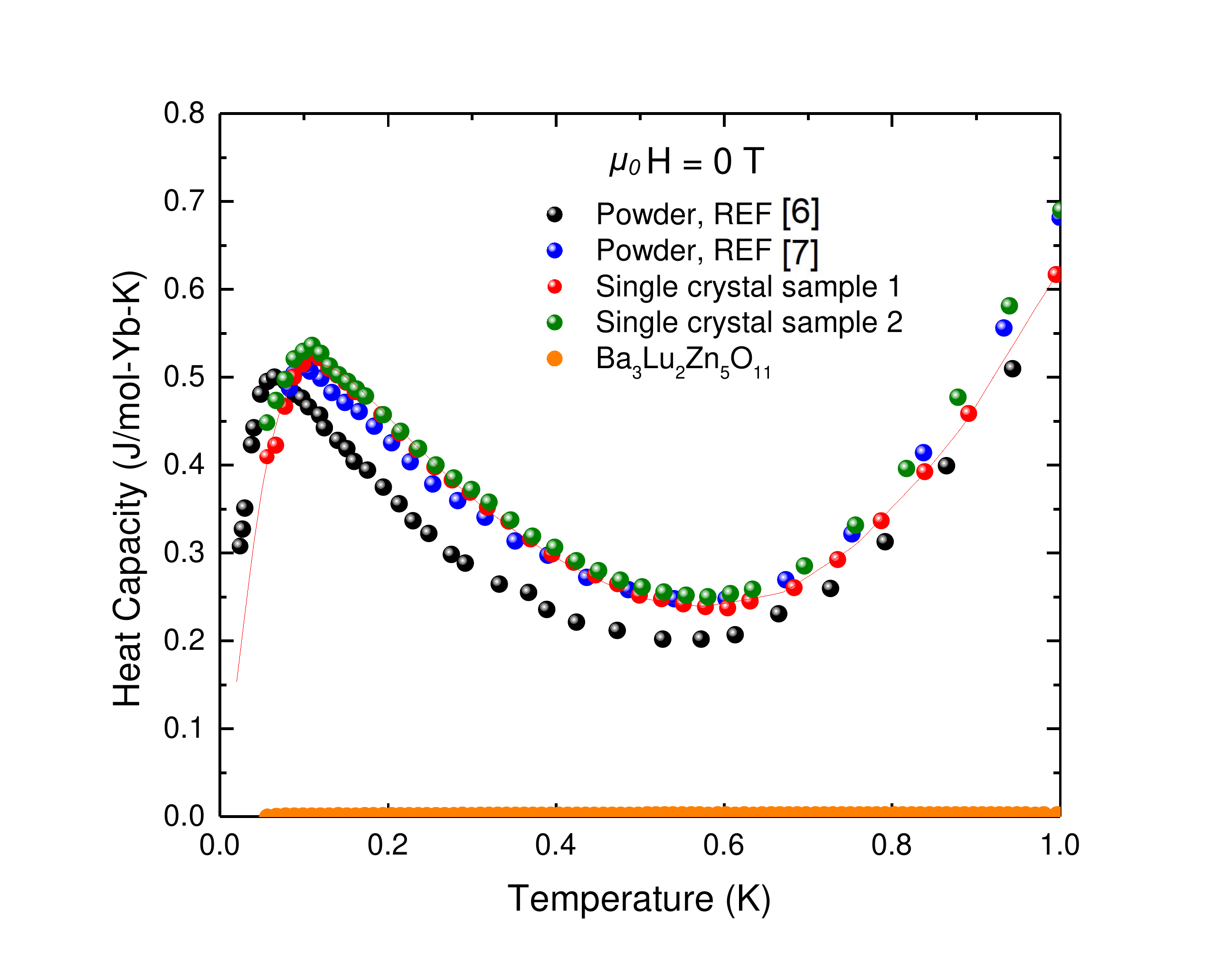}
	\caption{Low temperature heat capacity data of \byzo{} single crystal samples are shown. Two different single crystal samples with different growth conditions show the same behavior at low temperatures with a broad peak at 110 mK. The heat capacity data reported by Haku \textit{et al.}~\cite{HakuPRB2016} and Rau \textit{et al.}~\cite{RauJPCM_2018} are compared. The phonon contribution was measured using the iso-structural, non-magnetic compound \blzo{} and found almost zero below 1 K. Red solid line indicates the fit to the low temperature data as explained in the main text.}
	
	\label{fig:HC_supp}
\end{figure}

\subsection{ac-Susceptibility and Tunnel Diode Oscillator}
\label{sec:Susceptibility and Magnetization}

To probe the high-field magnetic properties of the \byzo{} crystals, we conducted the magnetic ac-susceptibility and tunnel diode oscillator (TDO) measurements at the DC Field Facility of the National High Magnetic Field Laboratory in Tallahassee, FL. Dilution refrigerator and $ ^{3} $He systems were used to cover the temperature range from 20 K down to  41 mK. The field dependent measurements up to 18 T were conducted using a superconducting magnet. The low field sweep rate of 0.1-0.3 T/min was used to minimize the magnetocaloric effect.

The TDO measurements \cite{CloverTDO2008,VanTDO1975} were carried out on bar-shaped ~\byzo{} single crystals of $\sim$ 2 mm in length and $\sim$ 1 mm in width and depth. The crystals were placed inside a detection coil, with the $[111]$ direction aligned with the coil axis. The coil and sample within form the inductive component of a LC circuit. The LC circuit, powered by a tunnel diode operating in its negative resistance region, was tuned to resonance at a frequency range between 10 and 50 MHz. The shift in the resonance frequency, which is related to the change in the sample magnetization, was then recorded. With this method, the changes in the magnetic moments can be measured to a very high precision $\sim$ $10^{-12}$ emu \cite{VanTDO1975,Shi2019}. 

The magnetic ac-susceptibility measurements were also conducted to complement the TDO measurements on \byzo{} single crystals. The samples have a typical dimension of $\sim$ 4 mm in length and $\sim$ 0.8 mm in width and depth. A standard two-coil set-up was used. Here, a primary coil was used to apply an oscillating ac-magnetic-field ($\sim$ 0.1 Oe - 10 Oe) to the sample, and a secondary coil was used to detect the inducted voltage. The measurements were also repeated for different frequencies: 87.1 Hz, 377.7 Hz, and 1616 Hz. However, We found no dependence on the frequency of the ac-magnetic-field.

\begin{figure}[b!]
	\centering
	\includegraphics[width=1\columnwidth]{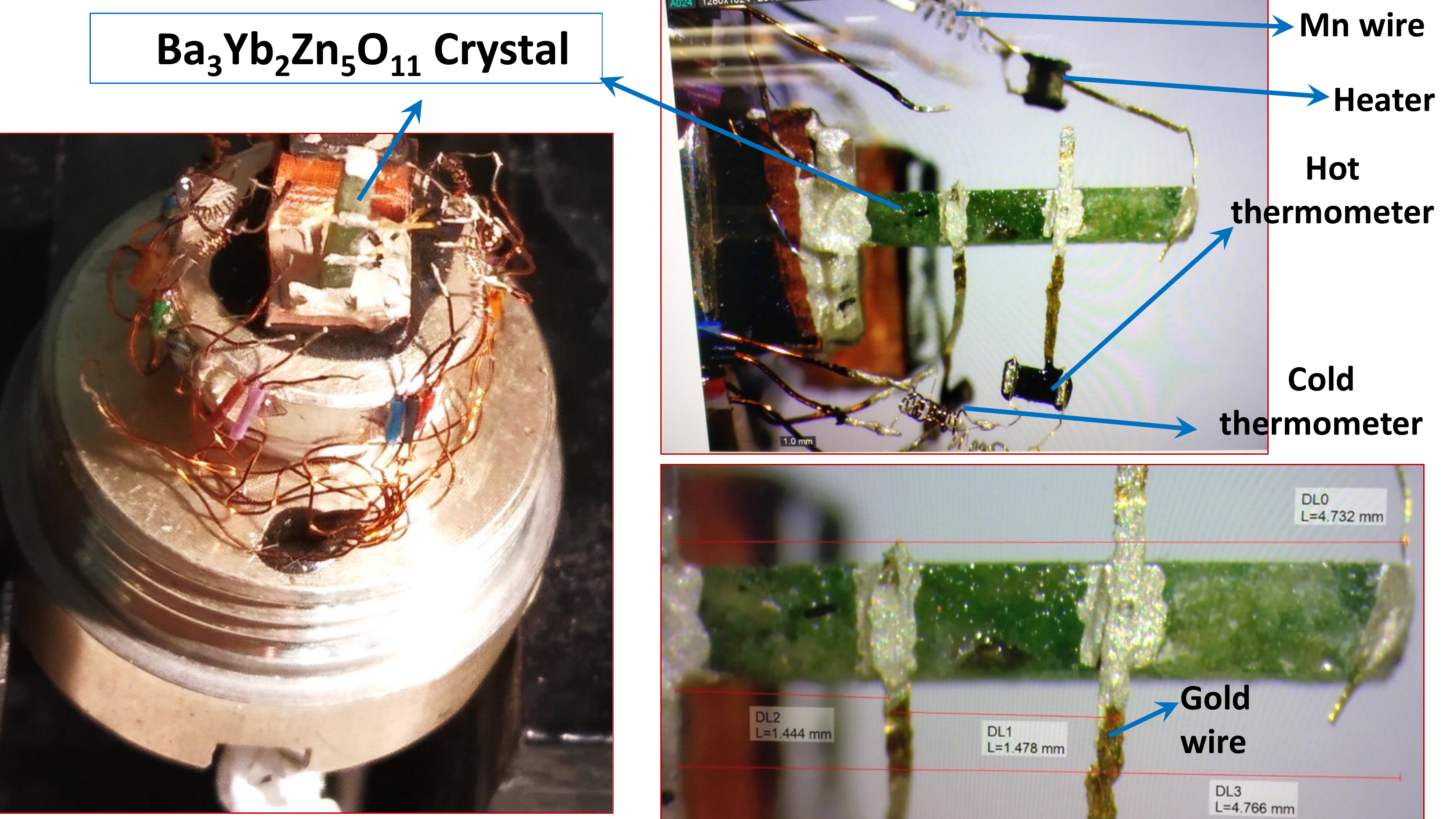}
	\caption{The \byzo{} crystal was glued with Dupont silver paint to a copper heat sink at the bottom of the vacuum cell, which was thermally anchored to the cold head of the cryostat. Using silver epoxy Dupont H20E, a heater was attached to the other end of the crystal so that a heat current could be applied. Two RuO$ _{2} $ sensors were attached to the side of the crystal for measurements of longitudinal temperature difference $ \nabla_{x}T $ along $[111]$.
	} 
	\label{fig:thermalcond}
\end{figure}

\subsection{Heat Capacity}
\label{sec:Specific Heat}

Low temperature heat capacity measurements were performed on single crystal samples of \byzo{} using a dilution refrigerator insert in a Quantum Design Physical Property Measurement System (QD PPMS) \footnote{The identification of any commercial product or trade name does not imply
endorsement or recommendation by the National Institute of Standards and Technology.}. Two single crystal samples grown with two different growth conditions were used (as explained above). Fig. \ref{fig:HC_supp} shows comparison plot of the measured heat capacity data on two different \byzo{} single crystals (sample 1 and sample 2) along with the previously reported powder data ($\mu_0 H =$ 0 T) \cite{RauJPCM_2018, HakuPRB2016}. The heat capacity as function of temperature ($0.5 \K< T <4 \K$) for several applied magnetic fields were also studied. The sample was mounted such that the magnetic field was along $[111]$. In order to measure phonon contributions, the iso-structural non-magnetic compound \blzo{} was also measured under zero field and subtracted from the total heat capacity data to calculate the magnetic heat capacity (C$_m$) of \byzo{}.

\subsection{Thermal Conductivity}
\label{sec:Thermal and thermal Hall conductivity measurements}

Low temperature thermal conductivity measurements were carried out on different single crystal samples of \byzo{} at the National High Magnetic Field Laboratory. The samples were cut into bar shaped crystals of $\sim$ 5.0 mm $\times$ 0.7 mm $\times$ 0.7 mm (length $\times$ width $\times$ depth) for the thermal conductivity ($\kappa_{xx}$) measurements. The samples were prepared such that the heat current density $Q$ (along the length of the sample) was along the $[111]$ direction. Magnetic field was applied both parallel and perpendicular to the $[111]$ direction, and no difference was observed. The repeated measurement of $\kappa_{xx}$ on different \byzo{} crystal is found reproducible. Fig. \ref{fig:thermalcond} shows a representative image of the mounted \byzo{} single crystals with cold and hot thermometers during thermal conductivity measurement. The steady-state method was used for the thermal conductivity measurements. The crystal was glued with Dupont silver paint to a copper heat sink at the bottom of the vacuum cell, which was thermally anchored to the cold head of the cryostat. Using silver epoxy Dupont H20E, a heater was attached to the other end of the crystal so that a heat current could be applied. Two RuO$ _{2} $ sensors were attached to the side of the crystal for measurements of thermal gradient ($ \nabla_{x}T $) along length. 

\subsection{Polarized Neutron Scattering}

Polarized neutron scattering experiments were performed at Multi Axis Chopper Spectrometer (MACS) at National Institute of Standards and Technology \cite{rodriguez2008macs,ye2013wide, fu2011wide, chen2016recent}. A single crystal sample of mass of about 1 g was mounted on a Cu sample holder in a dilution refrigerator. The sample was mounted with the $[{h+k},-{h}+k,-2k]$  scattering plane being horizontal and with the vertical guide field applied along the $[111]$ cubic direction. A fixed final neutron energy of $E_f = 4.2$ meV used and the measurements were performed at the energy transfer $E=0.5$ meV. Elastic energy resolution with this configuration was about 0.3 meV at elastic channel. Initial flipping ratio was about 25. The polarized neutron data were corrected using polarized-beam correction software (Pbcor) to account for the time-dependent neutron polarization (hence flipping ratio) and transmission due to the decay of the $^3$He polarization with relaxation times of a few hundred hours. Data were collected at $T=0.1$ K and $T=50$ K.

Figs.~\ref{fig:BYZO_Polarized_Figure}(a,b) show the polarized neutron scattering data measured at $T=0.1$ K and $E=0.5$ meV, after subtracting data at $T=50$ K and $E=0.5$ meV as the background. In our polarized setup, the neutron polarization at the sample is perpendicular to wave vector Q. The NSF signal is sensitive to the nuclear coherent scattering, the magnetic spin fluctuation perpendicular to the scattering plane, and 1/3 of nuclear spin incoherent scattering (if any). While the SF signal is sensitive to the magnetic spin fluctuation perpendicular to Q in the scattering plane and 2/3 of nuclear spin incoherent scattering \cite{moon1969polarization}. Therefore, in this case, the NSF and SF scattering data represent the spin fluctuations along $[111]$ cubic direction and spin fluctuations in the $[{h+k},-{h}+k,-2k]$  plane, respectively. Fig.~\ref{fig:BYZO_Polarized_Figure}(c,d) show the calculations of NSF and SF scattering using single tetrahedron theory outlined in the following section.

\section{Theoretical Model}

\label{sec:theory}

\begin{figure}[ht]
	\centering 
	\subfloat[\label{Fig_Energy_level}]{\includegraphics[width=0.45\columnwidth]{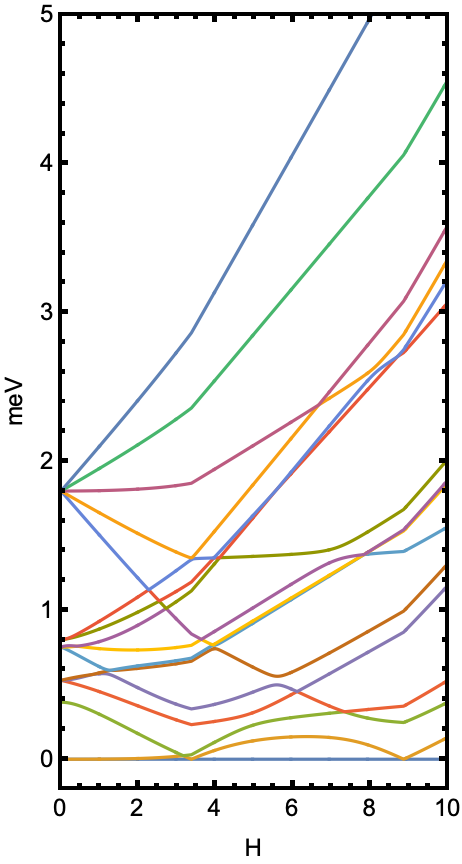}}\quad
	\subfloat[\label{Fig_Energy_level_para2}]{\includegraphics[width=0.45\columnwidth]{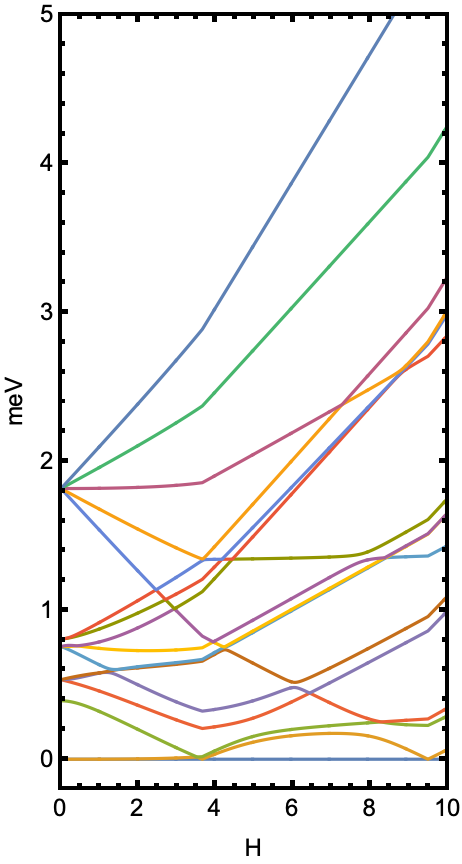}}
	\caption{(a) Energy levels of a single $A$-tetrahedron in magnetic field, using parameters from Eq.~\eqref{EQN_para_1}. (b) Energy levels   using parameters from Eq.~\eqref{EQN_para_2}.
	} 
	\label{fig:energylevels}
\end{figure}

In this section we describe the construction of the effective model used to produce the heat capacity curve in the main text, and related aspects of the theory. The breathing pyrochlore lattice has strong interactions on the $A$-tetrahedra, and much weaker interactions on the $B$-tetrahedra. Hence we first analyze the physics on a single $A$-tetrahedra, and then consider their interactions via bonds on the $B$-tetrahedra. The single tetrahedron theory has been proposed in previous works already \cite{HakuJPSJ2016,HakuPRB2016,RauPRL2016,RauJPCM_2018}. Its analysis can be done by directly diagonalizing the $2^4=16$ dimensional Hamiltonian of the 4 spins. The eigenstates can be  known directly up to numerical precision of the computer. The most general Hamiltonian constrained by symmetry is: 
\[
\mathcal{H}_\text{A-tetrahedron} = \sum_{ij} S_i ^\alpha J^{\alpha\beta}S^{\beta}
\]
where 
\[
\bm{J}_{01}  = 
\begin{pmatrix}
J_1 & J_3 & -J_4 \\
J_3 & J_1 & -J_4 \\
J_4 & J_4 & J_2 
\end{pmatrix},
\]
and other $J_{ij}$'s can be obtained by rotating the basis accordingly \cite{RossPRX2011}. 
From inelastic neutron scattering experiments, 
Rau \emph{et al.} estimated the exchange parameters to be \cite{RauPRL2016}:
\[
\begin{split}
\text{set 1: } &J_1 = 0.587 \text{ meV}\quad  J_2 = 0.573 \text{ meV}\quad  \\ 
&J_3 = -0.011 \text{ meV} \quad
J_4 = -0.117\text{ meV} \\  
&g_z = 3.07\quad  g_\pm = 2.36\ .
\end{split}
\label{EQN_para_1}
\] 
These parameters yield the 16 states' energy levels shown in Fig.~\ref{Fig_Energy_level}. Another set of estimated parameters are given in  Ref.\cite{RauJPCM_2018} published at a later date:
\[
\begin{split}
\text{set 2: }&J_1 = 0.592 \text{ meV}\quad   J_2 = 0.581 \text{ meV}\\  
&J_3 = -0.01\text{ meV}  \quad 
J_4 = -0.126 \text{ meV} \\ 
&g_z = 2.72\quad   g_\pm = 2.30.
\end{split}
\label{EQN_para_2}
\]
The energy levels are shown in Fig.~\ref{Fig_Energy_level_para2}. 

The biggest difference between the two sets of parameters is the value of $g_z$. Comparing the computed level crossing position in magnetic field and the TDO peaks measured in experiment, we find that parameter set 1 yields better agreement, and will use parameter set 1 from now on. At $\mu_0 H = 0$, the lowest two states are degenerate, and such degeneracy is robust for small variations of $\bm{J}_{ij}$. In the limit of anisotropic interactions being zero, the two degenerate states are the two dimer covering of the four sites. There are three such possible classical dimer covering states but one of them is linearly dependent of the the two others, when written as a quantum wave function. The three states and their relation is: 
\begin{widetext}
\begin{align}
    &|\text{singlet   (1,4) (2,3)}\rangle =\frac{1}{\sqrt{4}} \left ( |\u\u\d\d \rangle +  |\d\d\u\u \rangle -  |\u\d\u\d \rangle -  |\d\u\d\u \rangle \right) \\
    &|\text{singlet   (1,3) (2,4)}\rangle =\frac{1}{\sqrt{4}} \left (  |\u\u\d\d \rangle +  |\d\d\u\u \rangle -  |\u\d\d\u \rangle -  |\d\u\u\u \rangle \right)\\
    &|\text{singlet   (1,2) (3,4)}\rangle =\frac{1}{\sqrt{4}} \left (  |\u\d\u\d \rangle +  |\d\u\d\u \rangle -  |\u\d\d\u \rangle -  |\d\u\u\d \rangle \right)\\
    &|\text{singlet   (1,3) (2,4)}\rangle -|\text{singlet   (1,4) (2,3)}\rangle = |\text{singlet   (1,2) (3,4)}\rangle 
\end{align}
\end{widetext}
%
\begin{figure}[b!]
	\centering
\vspace{-5mm}	\includegraphics[width=\columnwidth]{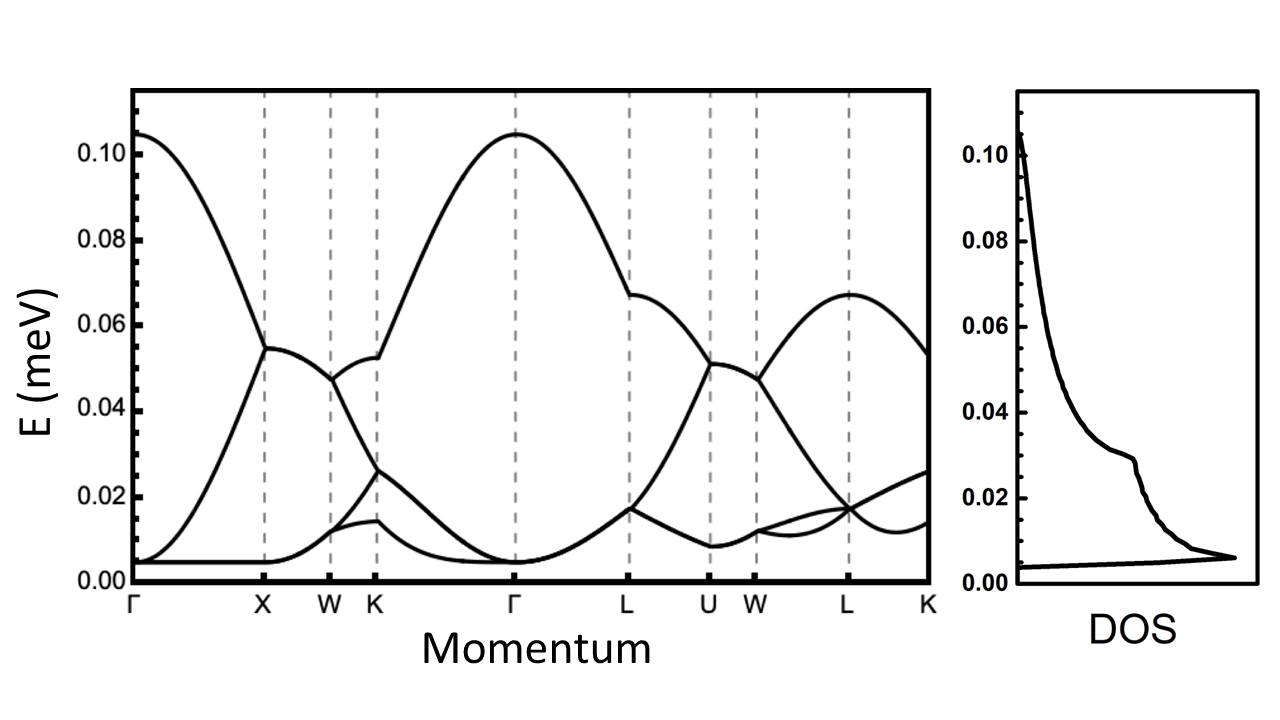}
	\caption{ Spin wave along high symmetry lines of FCC lattice, and the magnon density of states (DOS).}		
		\label{fig:FCC_SWDOS}
\end{figure}

With anisotropic interactions turned on, the two ground states stay degenerate, but the wave function changes. For small anisotropic interactions, the major contribution to the wave function is still the dimer covering states, but there is also  a fraction all-in and all-out wave function in the two states respectively, as well as other components. At finite external field $\mu_0 H$, the degeneracy of the two states is lifted. The energy gap is, however, very small for $\mu_0 H$ up to around $3$ T.

The uncorrelated $A$-tetrahedron theory can explain a lot of the experimental results \cite{RauJPCM_2018,RauPRL2016}, but  not  the specific heat peaks at low temperature of $0.1~\text{K} \sim 0.01~\text{meV}$. The broad peak at low temperature below the scale of the third eigenstate energy at $0.4~\text{meV}$ contributes entropy of about $\ln 2$ per $A$-tetrahedron, indicating that the two degenerate ground state must be lifted. Since no interactions within the $A$-tetrahedron can lift such a degeneracy, this is a strong evidence that inter-$A$-tetrahedron interactions, i.e., interaction via the $B$-tetrahedron bonds must exist. To demonstrate how this can provide an explanation for the specific heat, we  treat the two low-lying states as a pseudo spin-half on each $A$-tetrahedron, and consider their interactions via bonds on the $B$-tetrahedra. Shrinking every $A$-tetrahedron to a point and keeping the $B$-tetrahdra bonds, the lattice becomes a FCC lattice with nearest neighbor interactions (see Fig.~\ref{fig:Pyro_FCC_musr} in main text). The interactions between the pseudo spin-halves are the original interactions on the physical spins projected to the pseudo spin Hilbert space, represented by spin $\bm{s}_i$. The symmetry-allowed parameter range are vast, and current experimental data are not sufficient to narrow down the parameters. Here, we take the simplest approach by assuming the interaction to be:
\[
\mathcal{H}_\text{FCC} = \sum_{\langle ij \rangle}J_z s^z_i s^z_j + J_{xy}(s^x_i s^x_j + s^y_i s^y_j)
\]
is made of ferromagnetic $J_z s^z_i s^z_j$ interaction stabilizing a ferromagnetic ground state. In addition there is XY exchange $J_{xy}(s^x_i s^x_j + s^y_i s^y_j)$ that introduce dispersions of magnons (see Fig. \ref{fig:FCC_SWDOS}). We assume that the interactions between different bounds are identical in the global basis. The Hamiltonian can be diagonalized and various physical quantities including specific heat can be computed. 
We find the parameters:  
\[
J_z =-0.005~\text{meV},\ J_{xy}= 0.0125~\text{meV} 
\]
give rise to magnon dispersion relation that matches the specific heat measurements well. As mentioned, the parameter range is vast, and there is sure to be many other choices, but it is difficult to narrow down the parameters at present stage.

At $H\sim0.32$ T and $\mu_0 H\sim1.0 $ T, there are two peaks in ac-susceptibility,indicating potential phase transitions. For magnetic field in this range, the third eigenstate in the isolated $A$-tetrahedron theory is still far above the two lowest states. Hence, the phase transition must come from the varying inter-$A$-tetrahedron interactions, effectively described by the FCC lattice spin-halves. As $\mu_0 H$ increases, the wave function of the two low-lying states on the $A$-tetrahedron changes. As a consequence, we expect their effective interactions on the FCC lattice to change as well. Furthermore, since  the external field  $H$ is along $[111]$ direction, part of the symmetry of FCC lattice is broken. This further increases the allowed exchange terms on the FCC lattice model. Making a comprehensive account of phase diagram of this symmetry-broken FCC lattice is outside of the scope of this work. Given its high geometric frustration, this task is highly non-trivial and should be considered as one (or several) stand-alone project(s) by itself. Nevertheless, here we review some known results of FCC lattice spin models. Although the models may not directly apply to our case due to the fact that the pseudo spin-half behaves differently under symmetry compared to a physical spin-half, and the FCC lattice symmetry is further broken by the external field in $[111]$ direction, these known results are very helpful for us to have a qualitative understanding of the richness of the phase diagram. 

We note that with Heisenberg interactions on the nearest neighbors ($J_1$) and second neighbor ($J_2$), there are four phases on the phase diagram \cite{LefmannEPJB,SeehraPhysRevB1988}. One of them is ferromagnetic, when $J_1<0$ is  dominant. Depending on the values of $J_1,\ J_2$, there are three anti-ferromagnetic phases, each correspond to different ordering vectors $\bm{Q}$. We also note that the model with strong spin-orbital couplings allowing nearest neighbor anisotropic interactions has been studied in Ref.~\cite{CookPhysRevB2015}. The phase diagram contains various stripe phases, antiferromagnetic phases and spirals phases, with many phase transitions in-between. Beside these two studies of the general phase diagram, we also mention that on the breathing pyrochlore lattice model, and the perturbatively generated FCC lattice model,  it is possible to realize fracton-related physics. For example, Ref.~\cite{Yan2020PhysRevLett} proposed a breathing pyrochlore model on which a classical rank-2 U(1) spin liquid can be realized. This requires significantly larger Heisenberg interactions on the $B$-tetrahedron, when the perturbative treatment is not valid anymore. Han \emph {et al.}, also show that a quantum fracton model with emergent subsystem symmetry can exist in the perturbative regime \cite{HanPRB2022}.

\subsection{Polarized Neutron Scattering Calculation from Single Tetrahedron Theory}

The calculation of polarized inelastic neutron scattering from the single tetrahedron theory follows Ref.~\cite{RauPRL2016}.
For each single tetrahedron, there are 16 states labled by $\ket{n}$, $n=1,2,\dots, 16$. Their corresponding energy  $E_n$ and wave-function can be computed by exact diagonlization of the Hamiltonian.

The inelastic neutron scattering intensity is given by:
\[
I(\bm{Q}, \omega)=I_0 \frac{\left|\bm{k}^{\prime}\right|}{|\bm{k}|} \sum_{\alpha \beta}\left(\delta_{\alpha \beta}-\hat{Q}_\alpha \hat{Q}_\beta\right) F(|\bm{Q}|)^2 \mathcal{S}^{\alpha \beta}(\bm{Q}, \omega) .
\]
Here, $I_0$ is a normalization factor.
$\bm{k}$, $\bm{k}^{\prime}$ are the incoming and outgoing momenta of the neutron.
$F(|\bm{Q}|)$ is the atomic form factor for Yb$^{3+}$, which is assumed to be uniform because of the small range of $\bm{Q}$ measured in our experiment. 
The core quantity 
$\mathcal{S}^{\alpha \beta}(\bm{Q}, \omega)$ is the dynamical structure factor, given by:
\[
\mathcal{S}^{\alpha \beta}(\bm{Q}, \omega)=\sum_{n n^{\prime}} \frac{e^{-\beta E_n}}{Z}\left\langle n\left|\mu_{-\bm{Q}}^\alpha\right| n^{\prime}\right\rangle\left\langle n^{\prime}\left|\mu_{\bm{Q}}^\beta\right| n\right\rangle \delta\left(\omega-E_{n^{\prime}}+E_n\right).
\]
Here, $n, n'$ run  over all 16 states. $Z = \sum_n e^{-\beta E_n}$ is the partition function.
$\bm{\mu}_{\bm{Q}}$ is an operator for the neutron-spin coupling, defined as:
\[
\bm{\mu}_{\bm{Q}} \equiv \frac{1}{4} \sum_{i=1}^4 e^{-i \bm{Q} \cdot \bm{r}_i} \bm{\mu}_i.
\]
Here, $\bm{r}_i$ is the location of the four sites, and $\bm{\mu}_i$ is the effective moment of spin on site $i$ that couples to neutron:
\[
\bm{\mu}_i = \mu_B\left[g_{\pm}\left(\hat{\bm{x}}_i \tilde{S}_i^x+\hat{\bm{y}}_i \tilde{S}_i^y\right)+g_z \hat{\bm{z}}_i \tilde{S}_i^z\right].
\]
It is defined in local basis but need to be converted to global basis. In terms of the global basis $\hat{\bm{x}}, \hat{\bm{y}}, \hat{\bm{z}}$, the conversion is  : 
\[
\begin{array}{ll}
\hat{\bm{z}}_1=\frac{1}{\sqrt{3}}(+\hat{\bm{x}}+\hat{\bm{y}}+\hat{\bm{z}}), & \hat{\bm{x}}_1=\frac{1}{\sqrt{6}}(-2 \hat{\bm{x}}+\hat{\bm{y}}+\hat{\bm{z}}) \\
\hat{\bm{z}}_2=\frac{1}{\sqrt{3}}(+\hat{\bm{x}}-\hat{\bm{y}}-\hat{\bm{z}}), & \hat{\bm{x}}_2=\frac{1}{\sqrt{6}}(-2 \hat{\bm{x}}-\hat{\bm{y}}-\hat{\bm{z}}) \\
\hat{\bm{z}}_3=\frac{1}{\sqrt{3}}(-\hat{\bm{x}}+\hat{\bm{y}}-\hat{\bm{z}}), & \hat{\bm{x}}_3=\frac{1}{\sqrt{6}}(+2 \hat{\bm{x}}+\hat{\bm{y}}-\hat{\bm{z}}) \\
\hat{\bm{z}}_4=\frac{1}{\sqrt{3}}(-\hat{\bm{x}}-\hat{\bm{y}}+\hat{\bm{z}}), & \hat{\bm{x}}_4=\frac{1}{\sqrt{6}}(+2 \hat{\bm{x}}-\hat{\bm{y}}+\hat{\bm{z}})
\end{array}
\]
and $\hat{\bm{y}}_i=\hat{\bm{z}}_i \times \hat{\bm{x}}_i$.
Correspondingly, $\tilde{S}_i^\alpha$ are spin operators rotated from the global  to  local basis too.

\clearpage
\bibliography{main}